\documentclass[10pt]{llncs}
\usepackage{llncsdoc}
\usepackage{mathtools}
\usepackage{amssymb}
\usepackage{booktabs}
\usepackage{tikz-cd}
\usepackage{multirow}
\usepackage{subfigure}
\usepackage{tikz}
\usepackage{algorithm}
\usepackage{algorithmic}

\usetikzlibrary{positioning}

\begin{document}
\title{MetaGraph2Vec: Complex Semantic Path Augmented Heterogeneous Network Embedding}
\author{Daokun Zhang$^{1}$, Jie Yin$^{2}$, Xingquan Zhu$^{3}$, and Chengqi Zhang$^{1}$}
\institute{$^{1}$ Centre for Artificial Intelligence, FEIT, University of Technology Sydney, Australia\\
	\email{Daokun.Zhang@student.uts.edu.au, Chengqi.Zhang@uts.edu.au}\\
$^{2}$ Discipline of Business Analytics, The University of Sydney, Sydney, Australia\\
	\email{jie.yin@sydney.edu.au}\\
$^{3}$ Dept. of CEECS, Florida Atlantic University, USA\\
	\email{xqzhu@cse.fau.edu}}
\maketitle
\begin{abstract}
Network embedding in heterogeneous information networks (HINs) is a challenging task, due to complications of different node types and rich relationships between nodes. As a result, conventional network embedding techniques cannot work on such HINs. Recently, metapath-based approaches have been proposed to characterize relationships in HINs, but they are ineffective in capturing rich contexts and semantics between nodes for embedding learning, mainly because (1) metapath is a rather strict single path node-node relationship descriptor, which is unable to accommodate variance in relationships, and (2) only a small portion of paths can match the metapath, resulting in sparse context information for embedding learning. In this paper, we advocate a new metagraph concept to capture richer structural contexts and semantics between distant nodes. A metagraph contains multiple paths between nodes, each describing one type of relationships, so the augmentation of multiple metapaths provides an effective way to capture rich contexts and semantic relations between nodes. This greatly boosts the ability of metapath-based embedding techniques in handling very sparse HINs. We propose a new embedding learning algorithm, namely MetaGraph2Vec, which uses metagraph to guide the generation of random walks and to learn latent embeddings of multi-typed HIN nodes. Experimental results show that MetaGraph2Vec is able to outperform the state-of-the-art baselines in various heterogeneous network mining tasks such as node classification, node clustering, and similarity search. %


\end{abstract}
\section{Introduction}
Recent advances in storage and networking technologies have resulted in many applications with interconnected relationships between objects.
This has led to the forming of gigantic inter-related and multi-typed heterogeneous information networks (HINs) across a variety of domains, such as e-government, e-commerce, biology, social media, etc. HINs provide an effective graph model to characterize the diverse relationships among different types of nodes. Understanding the vast amount of semantic information modeled in HINs has received a lot of attention. In particular, the concept of metapaths~\cite{Sun:2011:pathsim}, which connect two nodes through a sequence of relations between node types, is widely used to exploit rich semantics in HINs. In the last few years, many metapath-based algorithms are proposed to carry out data mining tasks over HINs, including similarity search~\cite{Sun:2011:pathsim}, personalized recommendation~\cite{Jamali:2013:HeteroMF,Shi:2015:semantic}, and object clustering~\cite{Sun:2012:integrating}.   


Despite their great potential, data mining tasks in HINs often suffer from high complexity, because real-world HINs are very large and have very complex network structure. For example, when measuring metapath similarity between two distant nodes, all metapath instances need to be enumerated. This makes it very time-consuming to perform mining tasks, such as link prediction or similarity search, across the entire network. This inspires a lot of research interests in network embedding that aims to embed the network into a low-dimensional vector space, such that the proximity (or similarity) between nodes in the original network can be preserved. Analysis and search over large-scale HINs can then be applied in the embedding space, with the help of efficient indexing or parallelized algorithms designed for vector spaces.

\begin{figure}[!htbp]
	\begin{scriptsize}
		\centering
		\subfigure[Schema]{
			\label{fig1:schema}
			\begin{tikzpicture}  
			\node (V) at (0,0) [circle, draw, thick] {V};  
			\node[circle, draw, thick, above = 0.9 cm of V] (P) {P};
			\node[circle, draw, thick, above = 0.9 cm of P] (A) {A};
			\node[circle, below = 0.3 cm of V] (V0) {};
			\draw [->,thick] (A) -- node[midway,left] {$write$} (P);
			\draw [->,thick] (V) -- node[midway,left] {$publish$} (P);
			\draw [->,thick] (P) edge [loop, out=45, in=315, looseness=7] node[midway,right] {$cite$} (P);
			\end{tikzpicture}}
		\subfigure[Metapah and Metagraph]{
			\label{fig1:meta}
			\begin{tikzpicture}  
			\node (A1) at (0,0) [circle, draw, thick] {A};  
			\node[circle, draw, thick, right = 1 cm of A1] (P1) {P};
			\node[circle, right = 1 cm of P1] (N) {};
			\node[circle, draw, thick, right = 1 cm of N] (P2) {P};
			\node[circle, draw, thick, right = 1 cm of P2] (A2) {A};
			\node[circle, draw, thick, above =0.3 cm of N] (V1) {V};
			\node[circle, draw, thick, below = 0.3 cm of N] (A3) {A};
			\node[circle, below = 0.1 cm of A3] (V0) {};
			
			\node[circle, draw, thick, above = 0.5 cm of V1] (A4) {A};
			\node[circle, draw, thick, left = 1 cm of A4] (P3) {P};
			\node[circle, draw, thick, left = 1 cm of P3] (A5) {A};
			\node[circle, draw, thick, right = 1 cm of A4] (P4) {P};
			\node[circle, draw, thick, right = 1 cm of P4] (A6) {A};
			
			\node[circle, draw, thick, above = 0.5 cm of A4] (V2) {V};
			\node[circle, draw, thick, left = 1 cm of V2] (P5) {P};
			\node[circle, draw, thick, left = 1 cm of P5] (A7) {A};
			\node[circle, draw, thick, right = 1 cm of V2] (P6) {P};
			\node[circle, draw, thick, right = 1 cm of P6] (A8) {A};
			
			\node[left = 0.3cm of A1] (metagraph) {$\mathcal{G}:$};
			\node[left = 0.3cm of A5] (metapath1) {$\mathcal{P}_{2}:$};
			\node[left = 0.3cm of A7] (metapath2) {$\mathcal{P}_{1}:$};
			
			\draw [->,thick] (A1) -- node[midway,above] {\tiny$write$} (P1);
			\draw [->,thick] (P2) -- node[midway,above] {\tiny$write^{-1}$} (A2);
			\draw [->,thick] (P1) -- node[near end,left] {\tiny$publish^{-1}$} (V1);
			\draw [->,thick] (P1) -- node[near end,left] {\tiny$write^{-1}$} (A3);
			\draw [->,thick] (V1) -- node[near start,right] {\tiny$publish$} (P2);
			\draw [->,thick] (A3) -- node[near start,right] {\tiny$write$} (P2);
			
			\draw [->,thick] (A5) -- node[midway,above] {\tiny$write$} (P3);
			\draw [->,thick] (P3) -- node[midway,above] {\tiny$write^{-1}$} (A4);
			\draw [->,thick] (A4) -- node[midway,above] {\tiny$write$} (P4);
			\draw [->,thick] (P4) -- node[midway,above] {\tiny$write^{-1}$} (A6);
			
			\draw [->,thick] (A7) -- node[midway,above] {\tiny$write$} (P5);
			\draw [->,thick] (P5) -- node[midway,above] {\tiny$publish^{-1}$} (V2);
			\draw [->,thick] (V2) -- node[midway,above] {\tiny$publish$} (P6);
			\draw [->,thick] (P6) -- node[midway,above] {\tiny$write^{-1}$} (A8);
			
			\end{tikzpicture}}
		\caption{Schema, Metapath and Metagraph}
		\label{fig1} 
	\end{scriptsize}
\end{figure}
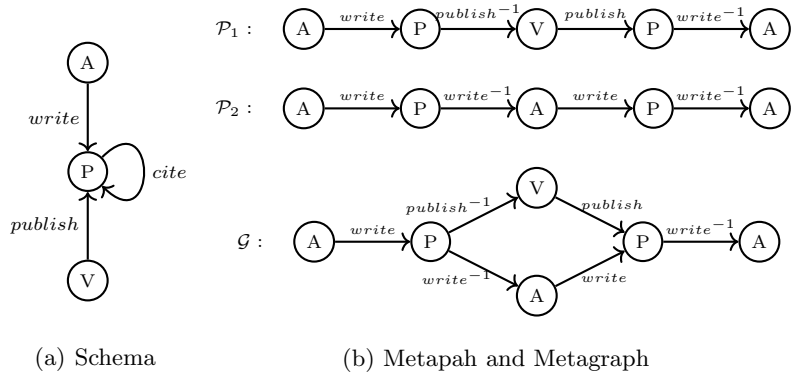

Conventional network embedding techniques~\cite{cao2015grarep,grover2016node2vec,perozzi2014deepwalk,tang2015line,wang2016structural,zhang2016homophily,zhang2017user,Zhang:2018:Survey}, however, focus on homogeneous networks, where all nodes and relations are considered to have a single type. Thus, they cannot handle the heterogeneity of node and relation types in HINs. Only very recently, metapath-based approaches~\cite{Chen:2017:task,Dong:2017:metapath2vec}, such as MetaPath2Vec~\cite{Dong:2017:metapath2vec}, are proposed to exploit specific metapaths as guidance to generate random walks and then to learn heterogeneous network embedding. For example, consider a DBLP bibliographic network, Fig.~\ref{fig1:schema} shows the HIN schema, which consists of three node types: Author (A), Paper (P) and Venue (V), and three edge types: an author writes a paper, a paper cites another paper, and a paper is published in a venue. The metapath $\mathcal{P}_{1}$: $A \rightarrow P \rightarrow V \rightarrow P \rightarrow A$ describes the relationship where both authors have papers published in the same venue, while $\mathcal{P}_{2}$: $A \rightarrow P \rightarrow A \rightarrow P \rightarrow A$ describes that two authors share the same co-author. If $\mathcal{P}_{1}$ is used by MetaPath2Vec to generate random walks, a possible random walk could be: $a_1 \rightarrow p_1 \rightarrow v_1 \rightarrow p_2 \rightarrow a_2$. Consider a window size of 2, authors $a_1$ and $a_2$ would share the same context node $v_1$, so they should be close to each other in the embedding space. This way, semantic similarity between nodes conveyed by metapaths is preserved. 

Due to difficulties in information access, however, real-world HINs often have sparse connections or many missing links. As a result, metapath-based algorithms may fail to capture latent semantics between distant nodes. As an example, consider the bibliographic network, where many papers may not have venue information, as they may be preprints submitted to upcoming venues or their venues are simply missing. The lack of paper-venue connection would result in many short random walks, failing to capture hidden semantic similarity between distant nodes. On the other hand, besides publishing papers on same venues, distant authors can also be connected by other types of relations, like sharing common co-authors or publishing papers with similar topics. Such information should be taken into account to augment metapath-based embedding techniques.


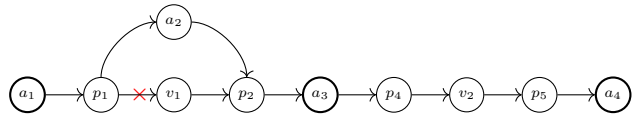
\begin{figure}[!htbp]
	\begin{scriptsize}
	\centering
	\begin{tikzpicture}  
	\node (a1) at (0,0) [thick, scale = 0.8, circle, draw] {$a_{1}$};  
	\node[scale = 0.8, circle, draw, right = 0.5cm of a1] (p1) {$p_{1}$};
	\node[scale = 0.8, circle, draw, right = 0.5cm of p1] (v1) {$v_{1}$};
	\node[scale = 1.2, right = of a1] (cross) {\color{red}$\mathbf{\times}$};
	\node[scale = 0.8, circle, draw, above = 0.5cm of v1] (a2) {$a_{2}$};
	\node[scale = 0.8, circle, draw, right = 0.5cm of v1] (p2) {$p_{2}$};
	\node[thick, scale = 0.8, circle, draw, right = 0.5cm of p2] (a3) {$a_{3}$};
	\node[scale = 0.8, circle, draw, right = 0.5cm of a3] (p4) {$p_{4}$};
	\node[scale = 0.8, circle, draw, right = 0.5cm of p4] (v2) {$v_{2}$};
	\node[scale = 0.8, circle, draw, right = 0.5cm of v2] (p5) {$p_{5}$};
	\node[thick, scale = 0.8, circle, draw, right = 0.5cm of p5] (a4) {$a_{4}$};
	
	\draw [->] (a1) -- (p1);
	\draw [->] (p1) -- (v1);
	\draw [->] (v1) -- (p2);
	\draw [->] (p2) -- (a3);
	\draw [->] (a3) -- (p4);
	\draw [->] (p4) -- (v2);
	\draw [->] (v2) -- (p5);
	\draw [->] (p5) -- (a4);
	
	\draw [->] (p1) edge [out=90, in=180, looseness=0.8]  (a2);
	\draw [->] (a2) edge [out=0, in=90, looseness=0.8]  (p2);
	\end{tikzpicture} 
	\caption{An example of random walk from $a_1$ to $a_4$ based on metagraph $\mathcal{G}$, which cannot be generated using metapaths $\mathcal{P}_1$ and $\mathcal{P}_2$. This justifies the ability of MetaGraph2Vec to provide richer structural contexts to measure semantic similarity between distant nodes.}
	\label{fig2:example} 
	\end{scriptsize}
\end{figure}

Inspired by this observation, we propose a new method for heterogeneous network embedding, called MetaGraph2Vec, that learns more informative embeddings by capturing richer semantic relations between distant nodes. The main idea is to use metagraph~\cite{Huang:2016:metastructure} to guide random walk generation in an HIN, which fully encodes latent semantic relations between distant nodes at the network level. Metagraph has its strength to describe complex relationships between nodes and to provide more flexible matching when generating random walks in an HIN. Fig.~\ref{fig1:meta} illustrates a metagraph $\mathcal{G}$, which describes that two authors are relevant if they have papers published in the same venue or they share the same co-authors. Metagraph $\mathcal{G}$ can be considered as a union of metapaths $\mathcal{P}_1$ and $\mathcal{P}_2$, but when generating random walks, it can provide a superset of random walks generated by both $\mathcal{P}_1$ and $\mathcal{P}_2$. Fig.~\ref{fig2:example} gives an example to illustrate the intuition behind. When one uses metapath $\mathcal{P}_1$ to guide random walks, if paper $p_1$ has no venue information, the random walk would stop at $p_1$ because the link from $p_1$ to $v_1$ is missing. This results in generating too many short random walks that cannot reveal semantic relation between authors $a_1$ and $a_3$. In contrast, when metagraph $\mathcal{G}$ is used as guidance, the random walk $a_1 \rightarrow p_1 \rightarrow a_2 \rightarrow p_2 \rightarrow a_3$, and $a_3 \rightarrow p_4 \rightarrow v_2 \rightarrow p_5 \rightarrow a_4$ is generated by taking the path en route $A$ and $V$ in $\mathcal{G}$, respectively. This testifies the ability of MetaGraph2Vec to provide richer structural contexts to measure semantic similarity between distant nodes, thereby enabling more informative network embedding. 

Based on this idea, in MetaGraph2Vec, we first propose metagraph guided random walks in HINs to generate heterogeneous neighborhoods that fully encode rich semantic relations between distant nodes. Second, we generalize the Skip-Gram model~\cite{mikolov2013distributed} to learn latent embeddings for multiple types of nodes. Finally, we develop a heterogeneous negative sampling based method that facilitates the efficient and accurate prediction of a node's heterogeneous neighborhood. MetaGraph2Vec has the advantage of offering more flexible ways to generate random walks in HINs so that richer structural contexts and semantics between nodes can be preserved in the embedding space.


The contributions of our paper are summarized as follows:
\begin{enumerate}
\item  We advocate a new \textit{metagraph} descriptor which augments metapaths for flexible and reliable relationship description in HINs. Our study investigates the ineffectiveness of existing metapath based node proximity in dealing with sparse HINs, and explains the advantage of metagraph based solutions. 
\item We propose a new network embedding method, called MetaGraph2Vec, that uses metagraph to capture richer structural contexts and semantics between distant nodes and to learn latent embeddings for multiple types of nodes in HINs. 
\item We demonstrate the effectiveness of our proposed method through various heterogeneous network mining tasks such as node classification, node clustering, and similarity search, outperforming the state-of-the-art.
\end{enumerate}

\section{Preliminaries and Problem Definition}
In this section, we formalize the problem of heterogeneous information network embedding and give some preliminary definitions.

\begin{definition}A \textbf{heterogeneous information network (HIN)} is defined as a directed graph $G=(V,E)$  with a node type mapping function $\phi:V\rightarrow\mathcal{L}$ and an edge type mapping function $\psi:E\rightarrow\mathcal{R}$. $T_{G}=(\mathcal{L},\mathcal{R})$ is the network schema that defines the node type set $\mathcal{L}$ with $\phi(v)\in\mathcal{L}$ for each node $v\in V$, and the allowable link types $\mathcal{R}$ with $\psi(e)\in\mathcal{R}$ for each edge $e\in E$.
\end{definition}

\begin{example}For a bibliographic HIN composed of authors, papers, and venues, Fig.~\ref{fig1:schema} defines its network schema. The network schema contains three node types, author (A), paper (P) and venue (V), and defines three allowable relations, $A\xrightarrow{write}P$, $P\xrightarrow{cite}P$ and $V\xrightarrow{publish}P$. Implicitly, the network schema also defines the reverse relations, i.e., $P\xrightarrow{write^{-1}}A$, $P\xrightarrow{cite^{-1}}P$ and $P\xrightarrow{publish^{-1}}V$.
\end{example}

\begin{definition}Given an HIN $G$, \textbf{heterogeneous network embedding} aims to learn a mapping function $\mathrm{\Phi}:V\rightarrow\mathbb{R}^{d}$ that embeds the network nodes $v\in V$ into a low-dimensional Euclidean space with $d\ll|V|$ and guarantees that nodes sharing similar semantics in $G$ have close low-dimensional representations $\mathrm{\Phi}(v)$. 
\end{definition}


\begin{definition} A \textbf{metagraph} is a directed acyclic graph (DAG) $\mathcal{G}=(N,M,n_{s},n_{t})$ defined on the given HIN schema $T_{G}=(\mathcal{L},\mathcal{R})$, which has only a single source node $n_{s}$ (\textit{i.e.}, with 0 in-degree) and a single target node $n_{t}$ (\textit{i.e.}, with 0 out-degree).  $N$ is the set of the occurrences of node types with $n\in\mathcal{L}$ for each $n\in N$. $M$ is the set of the occurrences of edge types with $m\in\mathcal{R}$ for each $m\in M$. \end{definition}

As metagraph $\mathcal{G}$ depicts complex composite relations between nodes of type $n_{s}$ and $n_{t}$, $N$ and $M$ may contain duplicate node and edge types. To clarify, we define the \textit{layer} of each node in $N$ as its topological order in $\mathcal{G}$ and denote the number of layers by $d_{\mathcal{G}}$. According to nodes' layer, we can partition $N$ into disjoint subsets $N[i]\ (1\leq i\leq d_{\mathcal{G}})$, which represents the set of nodes in layer $i$. Each $N[i]$ does not contain duplicate nodes. Now each element in $N$ and $M$ can be uniquely described as follows. For each $n$ in $N$, there exists a unique $i$ with $1\leq i\leq d_{\mathcal{G}}$ satisfying $n\in N[i]$ and we define the layer of node $n$ as $l(n)=i$. For each $m\in M$, there exist unique $i$ and $j$ with $1\leq i<j\leq d_{\mathcal{G}}$ satisfying $m\in N[i]\times N[j]$.

\begin{example}Given a bibliographic HIN $G$ and a network schema $T_{G}$ shown in Fig.~\ref{fig1:schema}, Fig.~\ref{fig1:meta} shows an example of metagraph $\mathcal{G}=(N,M,n_{s},n_{t})$ with $n_{s}=n_{t}=A$. There are $5$ layers in $\mathcal{G}$ and node set $N$ can be partitioned into 5 disjoint subsets, one for each layer, where $N[1]=\{A\}, N[2]=\{P\}, N[3]=\{A, V\}, N[4]=\{P\}, N[5]=\{A\}$. 
\end{example}

\begin{definition} For a metagraph $\mathcal{G}=(N,M,n_{s},n_{t})$ with $n_{s}=n_{t}$, its \textbf{recursive metagraph} $\mathcal{G}^{\infty}=(N^{\infty},M^{\infty},n^{\infty}_{s},n^{\infty}_{t})$ is a metagraph formed by tail-head concatenation of an arbitrary number of $\mathcal{G}$. $\mathcal{G}^{\infty}$ satisfies the following conditions:
\begin{enumerate}
	\item $N^{\infty}[i]=N[i]$ for $1\leq i< d_{\mathcal{G}}$, and $N^{\infty}[i]={N[i\ \mathrm{mod}\ d_{\mathcal{G}}+1]}$ for $i\geq d_{\mathcal{G}}$.
	\item For each $m\in N^{\infty}[i]\times N^{\infty}[j]$ with any $i$ and $j$, $m\in M^{\infty}$ if and only if one of the following two conditions is satisfied:
	\begin{enumerate}
		\item $1\leq i<j\leq d_{\mathcal{G}}$ and $m\in M\bigcap(N[i]\times N[j])$;
		\item $i\geq d_{\mathcal{G}}$, $1\leq j-i\leq d_{\mathcal{G}}$ and $m\in M\bigcap(N[i\mod d_{\mathcal{G}}+1]\times {N[j\mod d_{\mathcal{G}}+1]})$.
	\end{enumerate}
\end{enumerate}In the recursive metagraph $\mathcal{G}^{\infty}$, for each node $n\in N^{\infty}$, we define its layer as $l^{\infty}(n)$.
\end{definition}

\begin{definition}\label{randwalkseq}
Given an HIN $G$ and a metagraph $\mathcal{G}=(N,M,n_{s},n_{t})$ with $n_{s}=n_{t}$ defined on its network schema $T_{G}$, together with the corresponding recursive metagraph $\mathcal{G}^{\infty}=(N^{\infty},M^{\infty},n^{\infty}_{s},n^{\infty}_{t})$, we define the random walk node sequence constrained by metagraph $\mathcal{G}$ as $\mathcal{S}_{\mathcal{G}}=\{v_{1},v_{2},\cdots,v_{L}\}$ with length $L$ satisfying the following conditions: 
\begin{enumerate}
	\item For each $v_{i}\ (1\leq i\leq L)$ in $\mathcal{S}_{\mathcal{G}}$, $v_{i}\in V$ and for each $v_{i}\ (1< i\leq L)$ in $\mathcal{S}_{\mathcal{G}}$, $(v_{i-1},v_{i})\in E$. Namely, the sequence $\mathcal{S}_{\mathcal{G}}$ respects the network structure in $G$.
	\item $\phi(v_{1})=n_{s}$ and $l^{\infty}(\phi(v_{1}))=1$. Namely, the random walk starts from a node with type $n_{s}$.
	\item For each $v_{i}\ (1< i\leq L)$ in $\mathcal{S}_{\mathcal{G}}$, there exists a unique $j$ satisfying $(\phi(v_{i-1}),\phi(v_{i}))\in M^{\infty}\bigcap (N^{\infty}[l^{\infty}(\phi(v_{i-1}))]\times N^{\infty}[j])$ with $j>l^{\infty}(\phi(v_{i-1}))$, $\phi(v_{i})\in N^{\infty}[j]$ and $l^{\infty}(\phi(v_{i}))=j$. Namely, the random walk is constrained by the recursive metagraph $\mathcal{G}^{\infty}$.
\end{enumerate}
\end{definition}

\begin{example}
Given metagraph $\mathcal{G}$ in Fig.~\ref{fig1:meta}, a possible random walk is $a_1 \rightarrow p_1 \rightarrow v_1 \rightarrow p_2 \rightarrow a_2 \rightarrow p_3 \rightarrow a_3 \rightarrow p_4 \rightarrow a_5$. It describes that author $a_1$ and $a_2$ publish papers in the same venue $v_1$ and author $a_2$ and $a_5$ share the common co-author $a_3$. Compared with metapath $\mathcal{P}_1$ given in Fig.~\ref{fig1:meta}, metagraph $\mathcal{G}$ captures richer semantic relations between distant nodes. 
\end{example}

\section{Methodology}
In this section, we first present metagraph-guided random walk to generate heterogeneous neighborhood in an HIN, and then present the MetaGraph2Vec learning strategy to learn latent embeddings of multiple types of nodes. 

\subsection{MetaGraph Guided Random Walk}
In an HIN $G=(V,E)$, assuming a metagraph $\mathcal{G}=(N,M,n_{s},n_{t})$ with $n_{s}=n_{t}$ is given according to domain knowledge, we can get the corresponding recursive metagraph $\mathcal{G}^{\infty}=(N^{\infty},M^{\infty},n^{\infty}_{s},n^{\infty}_{t})$. After choosing a node of type $n_{s}$, we can start the metagraph guided random walk. We denote the transition probability guided by metagraph $\mathcal{G}$ at $i$th step as $\mathrm{Pr}(v_{i}|v_{i-1};\mathcal{G}^{\infty})$. According to Definition $\ref{randwalkseq}$, if $(v_{i-1},v_{i})\notin E$, or $(v_{i-1},v_{i})\in E$ but there is no link from node type $\phi(v_{i-1})$ at layer $l^{\infty}(\phi(v_{i-1}))$ to node type $\phi(v_{i})$ in the recursive metagraph $\mathcal{G}^{\infty}$, the transition probability $\mathrm{Pr}(v_{i}|v_{i-1};\mathcal{G}^{\infty})$ is $0$. The probability $\mathrm{Pr}(v_{i}|v_{i-1};\mathcal{G}^{\infty})$ for $v_{i}$ that satisfies the conditions of Definition $\ref{randwalkseq}$ is defined as
{\small\begin{equation}
\mathrm{Pr}(v_{i}|v_{i-1};\mathcal{G}^{\infty})=\frac{1}{T_{\mathcal{G}^{\infty}}(v_{i-1})}\times\frac{1}{|\{u|(v_{i-1},u)\in E, \phi(v_{i})=\phi(u)\}|}.
\end{equation}}Above, $T_{\mathcal{G}^{\infty}}(v_{i-1})$ is the number of edge types among the edges starting from $v_{i-1}$ that satisfy the constraints of the recursive metagraph $\mathcal{G}^{\infty}$, which is formalized as
{\footnotesize\begin{equation}
	T_{\mathcal{G}^{\infty}}(v_{i-1})={|\{j| (\phi(v_{i-1}),\phi(u))\in M^{\infty}\bigcap (N^{\infty}[l^{\infty}(\phi(v_{i-1}))]\times N^{\infty}[j]), (v_{i-1},u)\in E \}|},
	\end{equation}}and $|\{u|(v_{i-1},u)\in E, \phi(v_{i})=\phi(u)\}|$ is the number of $v_{i-1}$'s 1-hop forward neighbors sharing common node type with node $v_{i}$. 

At step $i$, the metagraph guided random walk works as follows. Among the edges starting from $v_{i-1}$, it firstly counts the number of edge types satisfying the constraints and randomly selects one qualified edge type. Then it randomly walks across one edge of the selected edge type to the next node. If there are no qualified edge types, the random walk would terminate. 

\subsection{MetaGraph2Vec Embedding Learning}

Given a metagraph guided random walk $\mathcal{S}_{\mathcal{G}}=\{v_{1},v_{2},\cdots,v_{L}\}$ with length $L$, the node embedding function $\mathrm{\Phi}(\cdot)$ is learned by maximizing the probability of the occurrence of $v_{i}$'s context nodes within $w$ window size conditioned on $\mathrm{\Phi}(v_{i})$:
{\small\begin{equation}
\min_{\mathrm{\Phi}}-\log\mathrm{Pr}(\{v_{i-w},\cdots,v_{i+w}\}\setminus v_{i}|\mathrm{\Phi}(v_{i})),
\end{equation}}where,
{\small\begin{equation}
\mathrm{Pr}(\{v_{i-w},\cdots,v_{i+w}\}\setminus v_{i}|\mathrm{\Phi}(v_{i}))=\prod_{j=i-w,j\neq i}^{i+w}\mathrm{Pr}(v_{j}|\mathrm{\Phi}(v_{i})).
\end{equation}}Following MetaPath2Vec~\cite{Dong:2017:metapath2vec}, the probability $\mathrm{Pr}(v_{j}|\mathrm{\Phi}(v_{i}))$ is modeled in two different ways:
\begin{enumerate}
	\item \textbf{Homogeneous Skip-Gram} that assumes the probability $\mathrm{Pr}(v_{j}|\mathrm{\Phi}(v_{i}))$ does not depend on the type of $v_{j}$, and thus models the probability $\mathrm{Pr}(v_{j}|\mathrm{\Phi}(v_{i}))$ directly by softmax:
	{\small\begin{equation}
	\mathrm{Pr}(v_{j}|\mathrm{\Phi}(v_{i}))=\frac{\exp(\mathrm{\Psi}(v_{j})\cdot\mathrm{\Phi}(v_{i}))}{\sum_{u\in V}\exp(\mathrm{\Psi}(u)\cdot\mathrm{\Phi}(v_{i}))}.
	\end{equation}}
	\item \textbf{Heterogeneous Skip-Gram} that assumes the probability  $\mathrm{Pr}(v_{j}|\mathrm{\Phi}(v_{i}))$ is related to the type of node $v_{j}$:
	{\small\begin{equation}
	\mathrm{Pr}(v_{j}|\mathrm{\Phi}(v_{i}))=\mathrm{Pr}(v_{j}|\mathrm{\Phi}(v_{i}),\phi(v_{j}))\mathrm{Pr}(\phi(v_{j})|\mathrm{\Phi}(v_{i})),
	\end{equation}}where the probability $\mathrm{Pr}(v_{j}|\mathrm{\Phi}(v_{i}),\phi(v_{j}))$ is modeled via softmax:
	{\small\begin{equation}
	\mathrm{Pr}(v_{j}|\mathrm{\Phi}(v_{i}),\phi(v_{j}))=\frac{\exp(\mathrm{\Psi}(v_{j})\cdot\mathrm{\Phi}(v_{i}))}{\sum_{u\in V, \phi(u)=\phi(v_{j})}\exp(\mathrm{\Psi}(u)\cdot\mathrm{\Phi}(v_{i}))}. 
	\end{equation}}
\end{enumerate}
To learn node embeddings, the MetaGraph2Vec algorithm first generates a set of metagraph guided random walks, and then counts the occurrence frequency $\mathbb{F}(v_{i},v_{j})$ of each node context pair $(v_{i},v_{j})$ within $w$ window size. After that, stochastic gradient descent is used to learn the parameters. At each iteration, a node context pair $(v_{i},v_{j})$ is sampled according to the distribution of $\mathbb{F}(v_{i},v_{j})$, and the gradients are updated to minimize the following objective,
{\small\begin{equation}\label{SGD_obj}
\mathcal{O}_{ij}=-\log\mathrm{Pr}(v_{j}|\mathrm{\Phi}(v_{i})).
\end{equation}}To speed up training, negative sampling is used to approximate the objective function:
{\small\begin{equation}\label{SGD_obj_1}
\mathcal{O}_{ij}=\log\sigma(\mathrm{\Psi}(v_{j})\cdot\mathrm{\Phi}(v_{i}))+\sum_{k=1}^{K}\log\sigma(-\mathrm{\Psi}(v_{N_{j,k}})\cdot\mathrm{\Phi}(v_{i})),
\end{equation}}where $\sigma(\cdot)$ is the sigmoid function, $v_{N_{j,k}}$ is the $k$th negative node sampled for node $v_{j}$ and $K$ is the number of negative samples.
For Homogeneous Skip-Gram, $v_{N_{j,k}}$ is sampled from all nodes in $V$; for Heterogeneous Skip-Gram, $v_{N_{j,k}}$ is sampled from nodes with type $\phi(v_{j})$. Formally, parameters $\mathrm{\Phi}$ and $\mathrm{\Psi}$ are updated as follows:
{\small\begin{equation}\label{update_para}
\begin{aligned}
\mathrm{\Phi}=\mathrm{\Phi}-\alpha\frac{\partial\mathcal{O}_{ij}}{\partial \mathrm{\Phi}};\ \ \ \ \mathrm{\Psi}=\mathrm{\Phi}-\alpha\frac{\partial\mathcal{O}_{ij}}{\partial \mathrm{\Psi}},
\end{aligned}
\end{equation}}where $\alpha$ is the learning rate.

The pseudo code of the MetaGraph2Vec algorithm is given in Algorithm~\ref{alg:metgraph2vec}.
\begin{algorithm}[htb]
	\caption{The MetaGraph2Vec Algorithm}
	\label{alg:metgraph2vec}
	\begin{algorithmic}[1]
		\REQUIRE ~~\\
		(1) A heterogeneous information network (HIN): $G=(V,E)$;\\
		(2) A metagraph: $\mathcal{G}=(N,M,n_{s},n_{t})$ with $n_{s}=n_{t}$;\\
		(3) Maximum number of iterations: $MaxIterations$;
		\ENSURE ~~\\
		Node embedding $\mathrm{\Phi}(\cdot)$ for each $v\in V$;
		\STATE $\mathbb{S}$ $\leftarrow$ generate a set of random walks according to $\mathcal{G}$;
		\STATE $\mathbb{F}(v_i,v_j)$ $\leftarrow$ count frequency of node context pairs ($v_{i},v_{j})$ in $\mathbb{S}$;
		\STATE $Iterations \leftarrow 0;$
		\REPEAT
		\STATE $(v_i,v_j) \leftarrow$ sample a node context pair according to the distribution of $\mathbb{F}(v_i,v_j)$;
		\STATE $(\mathrm{\Phi}, \mathrm{\Psi}) \leftarrow$ update parameters using $(v_i,v_j)$ and Eq.~(\ref{update_para});
		\STATE $Iterations \leftarrow Iterations+1$;
		\UNTIL {$convergence$ or $Iterations\ge MaxIterations$ }\label{code:iteration}
		\STATE \textbf{return} $\mathrm{\Phi}$;
	\end{algorithmic}
\end{algorithm} 

\section{Experiments}
In this section, we demonstrate the effectiveness of the proposed algorithms for heterogeneous network embedding via various network mining tasks, including node classification, node clustering, and similarity search.

\subsection{Experimental Settings}
For evaluation, we carry out experiments on the DBLP\footnote{https://aminer.org/citation (Version 3 is used)} bibliographic HIN, which is composed of papers, authors, venues, and their relationships. Based on paper's venues, we extract papers falling into four research areas: \textit{Database}, \textit{Data Mining}, \textit{Artificial Intelligence}, \textit{Computer Vision}, and preserve the associated authors and venues, together with their relations. To simulate the paper-venue sparsity, we randomly select 1/5 papers and remove their paper-venue relations. This results in a dataset that contains 70,910 papers, 67,950 authors, 97 venues, as well as 189,875 paper-author relations, 91,048 paper-paper relations and 56,728 venue-paper relations. 

To evaluate the quality of the learned embeddings, we carry out multi-class classification, clustering and similarity search on author embeddings. Metapaths and metagraph shown in Fig.~\ref{fig1:meta} are used to measure the proximity between authors. The author's ground true label is determined by research area of his/her major publications. 

We evaluate MetaGraph2Vec with Homogeneous Skip-Gram and its variant MetaGraph2Vec++ with Heterogeneous Skip-Gram. We compare their performance with the following state-of-the-art baseline methods:
\begin{itemize}
	\item DeepWalk~\cite{perozzi2014deepwalk}: It uses the uniform random walk that treats nodes of different types equally to generate random walks.
	\item LINE~\cite{tang2015line}: We use two versions of LINE, namely LINE\_1 and LINE\_2, which models the first order and second order proximity, respectively. Both neglect different node types and edge types.
	\item MetaPath2Vec and MetaPath2Vec++~\cite{Dong:2017:metapath2vec}: They are the state-of-the-art network embedding algorithms for HINs, with MetaPath2Vec++ being a variant of MetaPath2Vec that uses heterogeneous negative sampling. To demonstrate the strength of metagraph over metapath, we compare with different versions of the two algorithms: $\mathcal{P}_1$ MetaPath2Vec, $\mathcal{P}_2$ MetaPath2Vec and Mixed MetaPath2Vec, which uses $\mathcal{P}_1$ only, $\mathcal{P}_2$ only, or both, to guide random walks, as well as their counterparts, $\mathcal{P}_1$ MetaPath2Vec++, $\mathcal{P}_2$ MetaPath2Vec++, and Mixed MetaPath2Vec++.
\end{itemize}

For all random walk based algorithms, we start random walks with length $L=100$ at each author for $\gamma=80$ times, for efficiency reasons. For the mixed MetaPath2Vec methods, $\gamma/2=40$ random walks are generated by following metapaths $\mathcal{P}_1$ and $\mathcal{P}_2$, respectively. To improve the efficiency, we use our optimization strategy for all random walk based methods: After random walks are generated, we first count the co-occurrence frequencies of node context pairs using a window size $w=5$, and according to the frequency distribution, we then sample one node context pair to do stochastic gradient descent sequentially. For fair comparisons, the total number of samples (iterations) is set to 100 million, for both random walk based methods and LINE. For all methods, the dimension of learned node embeddings $d$ is set to $128$. 

\subsection{Node Classification Results}
We first carry out multi-class classification on the learned author embeddings to compare the performance of all algorithms. We vary the ratio of training data from 1\% to 9\%. For each training ratio, we randomly split training set and test set for 10 times and report the averaged accuracy.


\begin{table}[!htbp]
	\centering
	\caption{Multi-class author classification on DBLP}
	\label{author_classification}
	\begin{scriptsize}	
		\renewcommand{\arraystretch}{1}
		\setlength\tabcolsep{3pt}
		\begin{tabular}{lccccccccc}
			\toprule
			Method          &  1\%      	 &  2\%			&  3\% 			&  4\% 	 		&  5\% 			 &  6\%			&  7\% 		   & 8\% 		 & 9\%\\
			\midrule
			DeepWalk         & 82.39	&86.04	&87.16	&88.15	&89.10	&89.49	&90.02	&90.25	&90.56\\
			LINE\_1          &71.25	&79.25	&83.11	&85.60	&87.17	&88.29	&89.05	&89.45	&89.63\\
			LINE\_2          &75.70 &80.80	&82.49	&83.88	&84.83	&85.71	&86.58	&86.90	&86.93\\
			$\mathcal{P}_1$ MetaPath2Vec  & 83.24	&87.70	&88.42	&89.05	&89.26	&89.46	&89.51	&89.76	&89.69\\
			$\mathcal{P}_1$ MetaPath2Vec++& 82.14	&86.02	&87.04	&87.96	&88.47	&88.66	&88.90	&88.91	&89.02\\
			$\mathcal{P}_2$ MetaPath2Vec  & 49.59	&52.12	&53.76	&54.67	&55.68	&55.49	&55.83	&55.68	&56.07\\
			$\mathcal{P}_2$ MetaPath2Vec++& 50.31	&52.50	&53.72	&54.47	&55.53	&55.78	&56.30	&56.36	&57.02\\
			Mixed MetaPath2Vec & 83.86	&87.34	&88.37	&89.22	&89.70	&90.01	&90.37	&90.42	&90.71 \\
			Mixed MetaPath2Vec++& 83.08	&86.91	&88.13	&89.07	&89.69	&90.09	&90.58	&90.68	&90.87\\
			MetaGraph2Vec   &\textbf{85.76}	&\textbf{89.00}	&89.79	&90.55	&91.02	&91.30	&91.72	&92.13	&92.25\\
			MetaGraph2Vec++  & 85.20	&88.97	&\textbf{89.99}	&\textbf{90.78}	&\textbf{91.42}	&\textbf{91.65}	&\textbf{92.13}	&\textbf{92.42}	&\textbf{92.46}\\
			\bottomrule
		\end{tabular}
	\end{scriptsize}
\end{table}

Table \ref{author_classification} shows the multi-class author classification results in terms of accuracy (\%) for all algorithms, with the highest score highlighted by $\textbf{bold}$. Our MetaGraph2Vec and MetaGraph2vec++ algorithms achieve the best performance in all cases. The performance gain over metapath based algorithms proves the capacity of MetaGraph2Vec in capturing complex semantic relations between distant authors in sparse networks, and the effectiveness of the semantic similarity in learning informative node embeddings. By considering methpaths between different types of nodes, MetaPath2Vec can capture better proximity properties and learn better author embeddings than DeepWalk and LINE, which neglect different node types and edge types. 


\subsection{Node Clustering Results}
We also carry out node clustering experiments to compare different embedding algorithms. We take the learned author embeddings produced by different methods as input and adopt $K$-means to do clustering. With authors' labels as ground truth, we evaluate the quality of clustering using three metrics, including Accuracy, F score and NMI. From Table \ref{clustering}, we can see that MetaGraph2Vec and MetaGraph2Vec++ yield the best clustering results on all three metrics. 

\begin{table}[!htbp]
	\centering
	\caption{Author clustering on DBLP}
	\label{clustering}
		\begin{scriptsize}
		\begin{tabular}{lccc}
			\toprule
			Method           &  Accuracy(\%)  &   F(\%)   &  NMI(\%)\\
			\midrule
			DeepWalk         & 73.87	& 67.39	& 42.02\\
			LINE\_1          & 50.26	& 46.33 & 17.94\\
			LINE\_2          & 52.14 	& 45.89 & 19.55\\
			$\mathcal{P}_1$ MetaPath2Vec  & 69.39	& 63.05	& 41.72\\
			$\mathcal{P}_1$ MetaPath2Vec++ & 66.11	& 58.68	& 36.45\\
			$\mathcal{P}_2$ MetaPath2Vec  & 47.51	&43.30 & 6.17\\
			$\mathcal{P}_2$ MetaPath2Vec++& 47.65	& 41.48	& 6.56\\
			Mixed MetaPath2Vec  & 77.20	& 69.50	& 49.43\\
			Mixed MetaPath2Vec++& 72.36	& 65.09	& 42.40\\
			MetaGraph2Vec    & \textbf{78.00}	& \textbf{70.96}& \textbf{51.40}\\
			MetaGraph2Vec++  &77.48&70.69&50.60\\
			\bottomrule
		\end{tabular}
	\end{scriptsize}
\end{table}

\subsection{Node Similarity Search}
Experiments are also performed on similarity search to verify the ability of MetaGraph2Vec to capture author proximities in the embedding space. We randomly select 1,000 authors and rank their similar authors according to cosine similarity score. Table \ref{search} gives the averaged precision@100 and precision@500 for different embedding algorithms. As can be seen, our MetaGraph2Vec and MetaGraph2Vec++ achieve the best search precisions.

\begin{table}[!htbp]
	\centering
	\caption{Author similarity search on DBLP}
	\label{search}
	\begin{scriptsize}
		\begin{tabular}{lccc}
			\toprule
			Methods           &  Precision$@100$ (\%)  &   Precision$@500$ (\%) \\
			\midrule
			DeepWalk         & 91.65	& 91.44 \\
			LINE\_1          & 91.18	& 89.88 \\
			LINE\_2          & 91.92 	& 91.38	\\
			$\mathcal{P}_1$ MetaPath2Vec  & 88.21	& 88.64	\\
			$\mathcal{P}_1$ MetaPath2Vec++& 88.68	& 88.58	\\
			$\mathcal{P}_2$ MetaPath2Vec  & 53.98	& 44.11	\\
			$\mathcal{P}_2$ MetaPath2Vec++& 53.39	& 44.11	\\
			Mixed MetaPath2Vec  & 90.94	& 90.27	\\
			Mixed MetaPath2Vec++& 91.49	& 90.69	\\
			MetaGraph2Vec    & 92.50& \textbf{92.17}	\\
			MetaGraph2Vec++  & \textbf{92.59}     &91.92\\
			\bottomrule
		\end{tabular}
	\end{scriptsize}
\end{table}

\subsection{Parameter Sensitivity}
\begin{figure}[!htbp]
	\centering
	\subfigure[$\gamma$]{
		\label{fig:parameter:subfig:gamma}
		\includegraphics[width=1.5in]{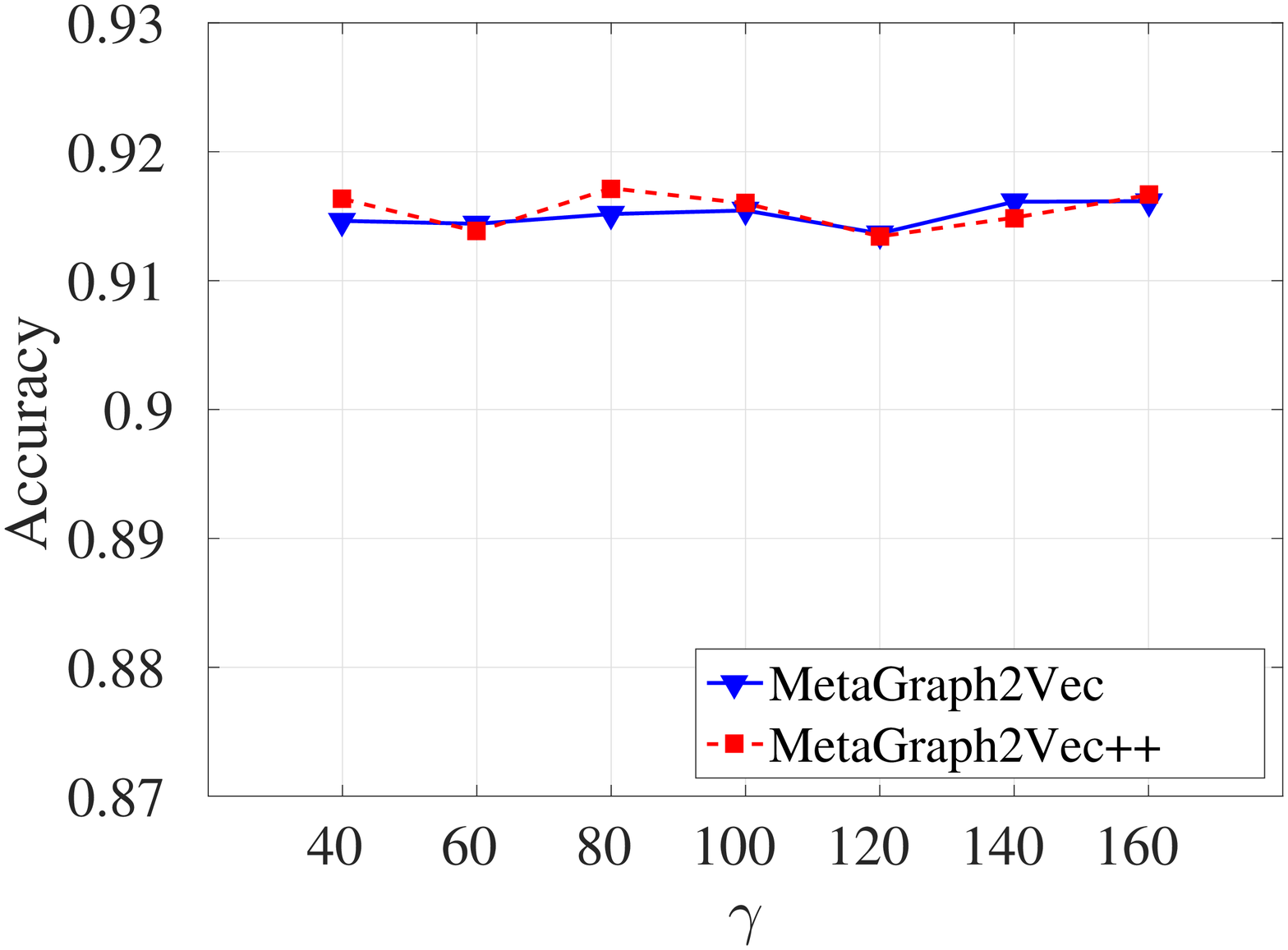}}
	\subfigure[$w$]{
		\label{fig:parameter:subfig:w}
		\includegraphics[width=1.5in]{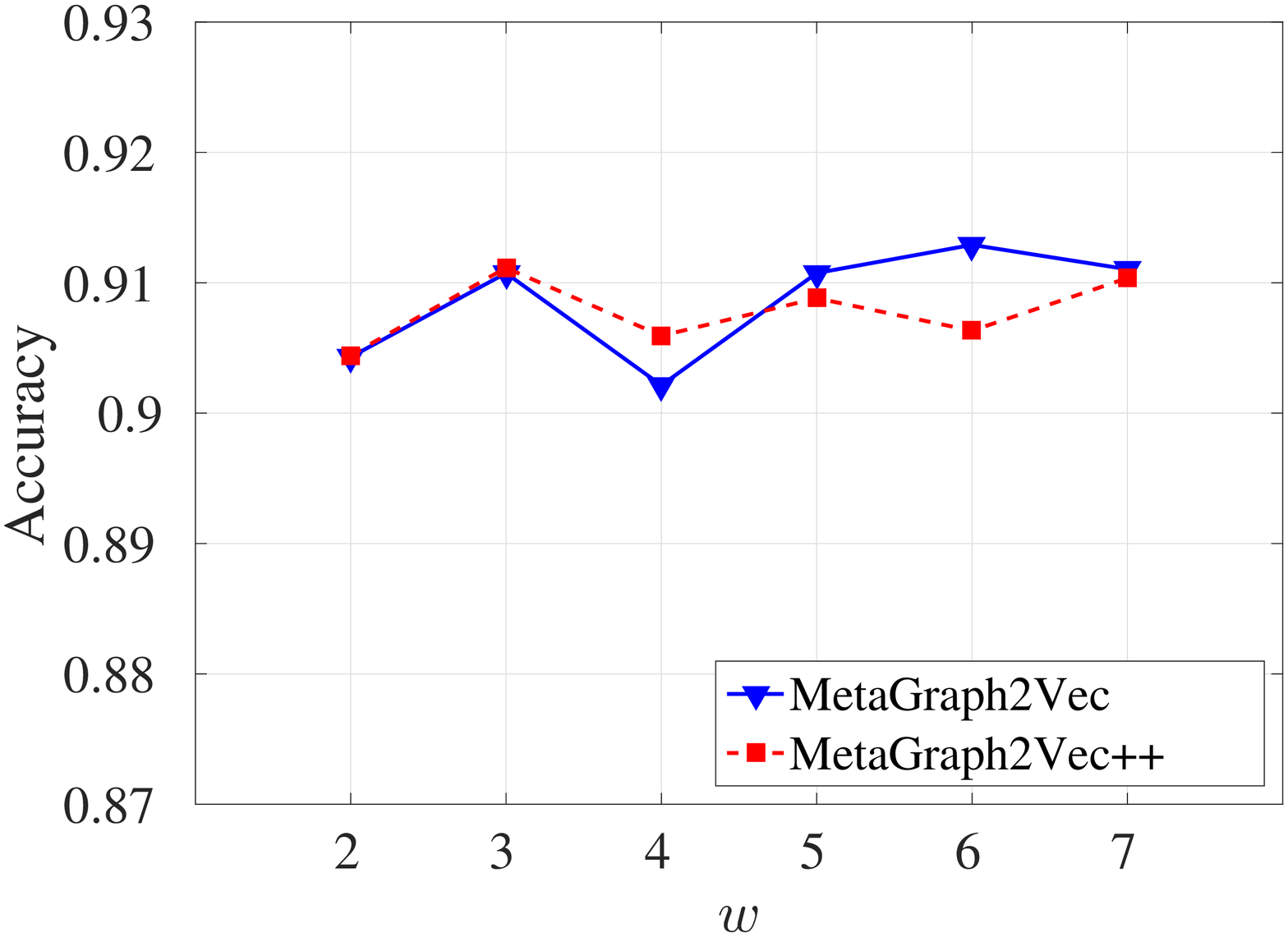}}
	\subfigure[$d$]{
		\label{fig:parameter:subfig:d}
		\includegraphics[width=1.5in]{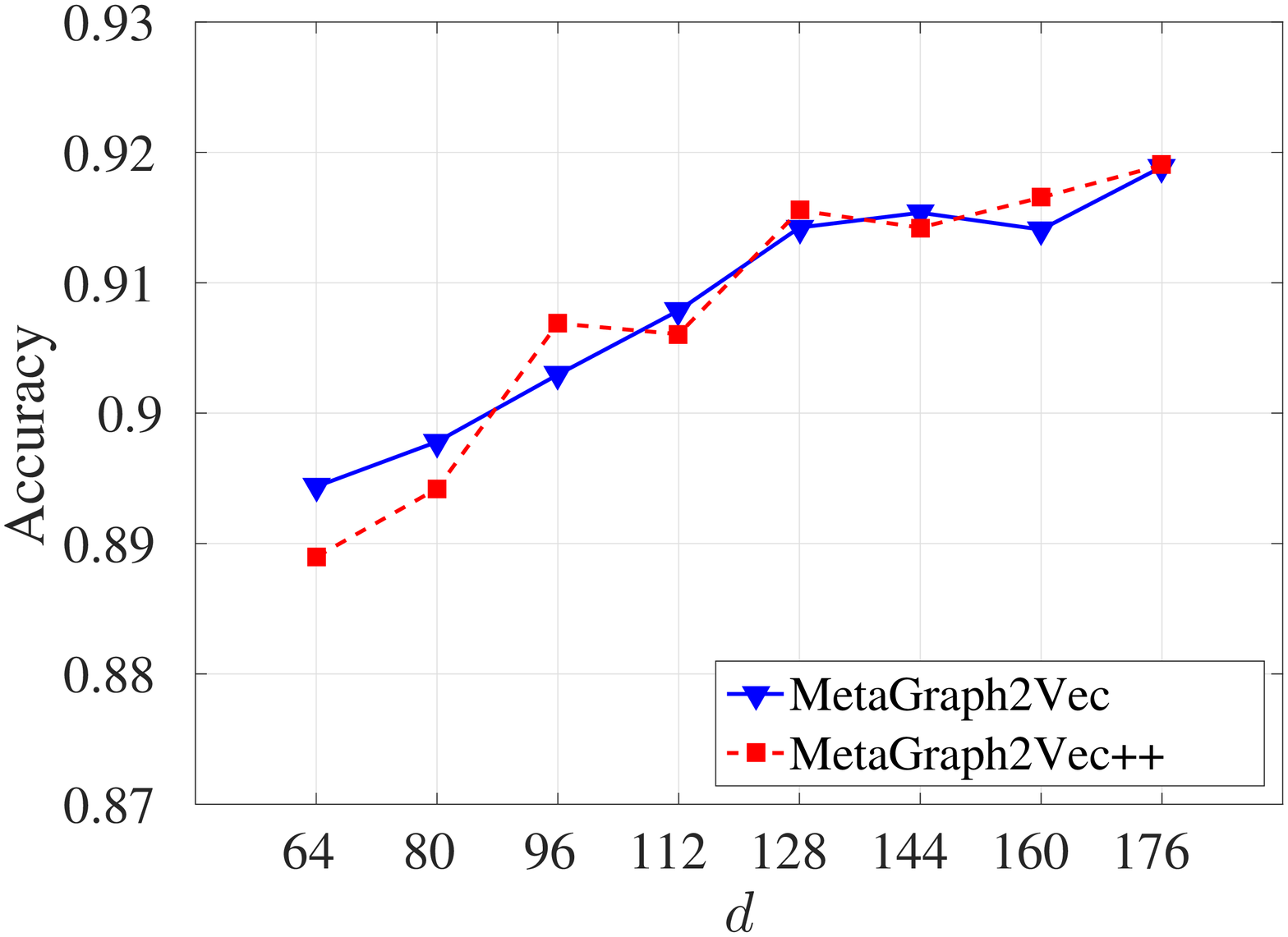}}
	\caption{The effect of parameters $\gamma$, $w$, and $d$ on node classification performance}
	\label{fig:parameter} 
\end{figure}

We further analyze the sensitivity of MetaGraph2vec and MetaGraph2Vec++ to three parameters: (1) $\gamma$: the number of metagraph guided random walks starting from each author; (2) $w$: the window size used for collecting node context pairs; (3) $d$: the dimension of learned embeddings. Fig.~\ref{fig:parameter} shows node classification performance with 5\% training ratio by varying the values of these parameters. We can see that, as the dimension of learned embeddings $d$ increases, MetaGraph2Vec and MetaGraph2Vec++ gradually perform better and then stay at a stable level. Yet, both algorithms are not very sensitive to the the number of random walks and window size. 

\section{Conclusions and Future Work}
This paper studied network embedding learning for heterogeneous information networks. We analyzed the ineffectiveness of existing \textit{metapath} based approaches in handling sparse HINs, mainly because metapath is too strict for capturing relationships in HINs. Accordingly, we proposed a new \textit{metagraph} relationship descriptor which augments metapaths for flexible and reliable relationship description in HINs. By using metagraph to guide the generation of random walks, our new proposed algorithm, MetaGraph2Vec, can capture rich context and semantic information between different types of nodes in the network. The main contribution of this work, compared to the existing research in the field, is twofold: (1) a new metagraph guided random walk approach to capturing rich contexts and semantics between nodes in HINs, and (2) a new network embedding algorithm for very sparse HINs, outperforming the state-of-the-art. 

In the future, we will study automatic methods for efficiently learning metagraph structures from HINs and assess the contributions of different metagraphs to network embedding. We will also evaluate the performance of MetaGraph2Vec on other types of HINs, such as heterogeneous biological networks and social networks, for producing informative node embeddings.\\

\noindent\textbf{Acknowledgments}. This work is partially supported by the Australian Research Council (ARC) under discovery grant DP140100545, and by the Program for Professor of Special Appointment (Eastern Scholar) at Shanghai Institutions of Higher Learning. Daokun Zhang is supported by China Scholarship Council (CSC) with No. 201506300082 and a supplementary postgraduate scholarship from CSIRO.

\bibliographystyle{plain}
\bibliography{hin,NRL-bibliography}

\end{document}